\documentstyle[12pt]{article}
\begin{document}
\def\F{{\cal F}}
\def\R{{{\rm I} \! {\rm R}}}
\vspace*{2cm}
\begin{center}
 \Huge\bf
Time machine as four-dimensional wormhole

\vspace*{0.25in}

\large

Alexandr K.\ Guts
\vspace*{0.15in}

\normalsize

Department of Mathematics, Omsk State University \\
644077 Omsk-77 RUSSIA
\\
\vspace*{0.5cm}
E-mail: guts@univer.omsk.su  \\
\vspace*{0.5cm}
December 25, 1996\\
\vspace{.5in}
ABSTRACT
\end{center}
The following mechanism of action of Time machine is considered.
Let space-time $<V^4, g_{ik}>$ be a leaf of a foliation $\F$ of
codimension 1 in
5-dimensional Lorentz manifold $<V^5, G_{AB}>$. If the
Godbillon-Vey class $GV(\F) \neq 0$ then the foliation $\F$ has resilient
leaves. Let $V^4$ be a resilient leaf.
Hence there exists an arbitrarily small neighborhood $U_a \subset V^5$
of the event $a \in V^4$
such  that $U_a \cap V^4$ consists of at least two connected components
$U_a^1$ and $U_a^2$.

Remove the four-dimensional balls
$B_a\subset U_a^1, B_b\subset U_a^2$, where an event $b\in U_a^2$,
and join the boundaries
of formed two holes by means of 4-dimensional cylinder. As result we
have a four-dimensional wormhole $C$, which is a Time machine if
$b$ belongs
to the past of
event $a$. The past of $a$ is lying arbitrarily nearly. The distant Past
is more accessible than the near Past.
It seems that real global space-time
$V^4$ is a resilient one, i.e. is a resilient leaf of some foliation $\F$.

It follows from the conformal Kaluza-Klein theory that the movement
to the Past
through four-dimensional wormhole $C$ along geodesic with respect
to metric $G_{AB}$
requires for time machine of large energy and electric
charge.

\newpage

\setcounter{page}{1}


We have a Time machine in a space-time domain $D$, when a smooth
closed time-like curve exists in this domain.
In the papers \cite{T1, T2} the case of creation of Time machine from the
three-dimensional wormhole (3-wormhole) by means of kinematic
procedures with one mouth
of wormhole is considered. The author agrees with opinion of
M.~Ju.~Konstantinov
\cite{K1, K2} that this assertion is erroneously because contradict to
Principle of equivalence and "in according with theorems about global
hyperbolicity and
Cauchy problem the models with causality
violation could not be considered as the result of dynamical evolution
of
some initial space-like configuration and must be considered as a
solution
of some boundary problem" \cite{K2}.  But Time machine exists if solution
of the Einstein's equations with 3-wormhole contains a closed
time-like curve,
i.e. if under creation of 3-wormhole simultaneously
the closed time-like curve is created.

In this paper the different principle of action of the
Time machine is discused.
 The closed time-like curve
can be constructed by means of
 attaching of four-dimensional handle ($4$-wormhole).

\section{Conditions of existence of Time machine in space-time
with Euclidean topology}

     Let $g_{ik} \ (i,k=0,...,3)$ be a constant gravitational
field in domain $D$, which is homeomorphic to Euclidean space $\R^4$
 and let this field is created by dust. Suppose that $g_{00}  >0$
and what is more  $ g_{00}= const=1$. The last relation can be reached
by means of transformation
$$
\bar x^0=\int\sqrt{g_{00}}dx^0,\qquad
\bar x^\alpha = x^\alpha, \ \ \  (\alpha=1,2,3).
$$
Assume that closed time-like curve $L$ is analytic Jordan
curve and it lies on 2-dimensional oriented surface $F\subset D$.
 And what is more suppose that $L$ is border of $F$.

It follows from formula \cite[p.357]{La} that
$$
g_{\alpha\beta}f_{\alpha\beta}f^{\alpha\beta} =\frac{16\pi G\rho}{c^2}
\frac{1+\frac{v^2}{c^2}}{1-\frac{v^2}{c^2}}, \eqno(1.1)
$$
where
$$
f_{\alpha\beta} =\frac{\partial}{\partial x^{\alpha}}\left(
\frac{g_{0\beta}}{g_{00}}\right) -
\frac{\partial}{\partial x^{\beta}}\left(
\frac{g_{0\alpha}}{g_{00}}\right),
$$
$\rho$ is density of dust, $v$ is velocity of dust.

The formula (1.1) can be rewrite in the form
$$
\Delta^3_3(f_{12})^2+ \Delta^2_2(f_{13})^2+\Delta^1_1(f_{23})^2+
2\Delta^3_2f_{12}f_{13}+2\Delta^3_1f_{12}f_{23}+
$$
$$
+2\Delta^2_1f_{13}f_{23}=
\frac{16\pi G\rho}{g_{00}c^2}
\frac{1+\frac{v^2}{c^2}}{1-\frac{v^2}{c^2}}, \eqno(1.2)
$$
where $\Delta^\alpha_\beta$ is minor of 2nd order of matrix
$\mid\mid -g^{\alpha\beta}\mid\mid$ that is got by means of
crossing out of line with number $\alpha$ and column with
number $\beta$.

Further we consider only such gravitational fields for which
either all
$$
\Delta^\alpha_\beta \geq 0,    \eqno(1.3)
$$
or, if $\Delta^\alpha_\beta <0$ then in (1.2) the summand
containing this $\Delta^\alpha_\beta$ is equal to zero.
The condition (1.3) is true for the G\"odel's solution, and
the second, for example, for metric
$$
ds^2=\frac{1}{2}(dx^0)^2+2\Omega x^2dx^0dx^1-\Omega x^1dx^0dx^2+
\left(\Omega^2(x^2)^2-\frac{1}{2}\right)(dx^1)^2-
$$
$$
-2\Omega^2x^1x^2dx^1dx^2+\left(\Omega^2(x^1)^2-\frac{1}{2}
\right)(dx^2)^2 +
A(dx^3)^2,
$$
which contains  smooth closed time-like curve of the form
$x^1=a\sin \mu, \ x^2=a\cos \mu, \ x^0, x^3=const$.

If $f_{\alpha\beta} <0 \ (\alpha<\beta)$ then we make the change
$f_{\alpha\beta}=-f_{\beta\alpha}$ and get that all
$f_{\alpha\beta}$
in (1.2) will be non-negative.

Let
$$
\Delta=\min\limits_{\alpha,\beta}\inf\limits_D\Delta^\alpha_\beta>0,
$$
where we considered only non-zero minors. It follows from (1.2)
that
$$
(f_{12}+f_{13}+f_{23})^2\Delta \leq
\frac{16\pi G\rho}{g_{00}c^2}
\frac{1+\frac{v^2}{c^2}}{1-\frac{v^2}{c^2}}. \eqno(1.4)
$$
Now we are able to calculate the chronometrical invariant time
$t(L)$ of passage of curve $L$ \cite{GD, GuDep}.
In the constant gravitational field
$f_{0i}=0$. By using this equality and the Stokes's formula me get
$$
t(L)=\frac{1}{c} \int\limits_{L}
\frac{ g_{0i}dx^i }{ \sqrt{g_{00}}}=
$$
$$
=\frac{ \sqrt{g_{00}} }{c} \int\hspace*{-2mm}\int
\limits_{\hspace*{-3mm}F}f_{12}dx^1dx^2
+f_{13}dx^1dx^3+f_{23}dx^2dx^3 =
$$
$$
=\frac{\sqrt{g_{00}}}{c}\int\hspace*{-2mm}\int
\limits_{\hspace*{-3mm}F}\left(f_{12}\cos\theta_1+
f_{13}\cos\theta_2+f_{23}\cos\theta_3\right)dS.
$$
By using (1.4) we  have further
$$
t(L) \leq \frac{\sqrt{g_{00}}}{c} \int\hspace*{-2mm}\int
\limits_{\hspace*{-3mm}F}
(f_{12}+f_{13}+f_{23})dS \leq
$$
$$
\leq \frac{4\sqrt{\pi G}}{c^2}\int\hspace*{-2mm}\int
\limits_{\hspace*{-3mm}F}\left(
\frac{\rho}{\Delta}\frac{1+\frac{v^2}{c^2}}{1-\frac{v^2}{c^2}}
\right)^\frac{1}{2}dS.
$$
Suppose that density $\rho$ and velocity $v$ of dust are constant
in domain $D$. Then
$$
t(L)\leq \frac{4}{c^2}
\left(
\frac{\pi G\rho}{\Delta}\frac{ 1+\frac{v^2}{c^2} }{ 1-\frac{v^2}{c^2} }
\right)^\frac{1}{2}\sigma(F),\eqno(1.5)
$$
where
$$
\sigma(F)= \int\hspace*{-2mm}\int\limits_{\hspace*{-3mm}F}dS
$$
is "Euclidean" area of surface $F$.

    If $s(L)$ is proper time of $L$ then
$$
s(L)=\int\limits_{L}\left(1-\frac{V^2}{c^2}\right)^\frac{1}{2}dt,
$$
where $V$ is velocity of Time machine with world line $L$,
$$
V^2=\gamma_{\alpha\beta}V^\alpha V^\beta, \qquad V^\alpha
=\frac{dx^\alpha}{dt},
$$
$$
\gamma_{\alpha\beta}=-g_{\alpha\beta}+
\frac{g_{0\alpha}g_{0\beta} }{g_{00}}.
$$
Hence proper time of return to the past can be less than any
positive small number. But
then the velocity of Time machine must tend to the velocity of light.

The formula (1.5) generalizes
(Demidov V.V. Master's degree work, Omsk State University, 1986. --
see \cite{GD, GuDep})
the analogeous estimate
$$
t(L)\sim\frac{\sqrt{8\pi G\rho}}{c^2}\sigma(F) \eqno(1.6)
$$
from \cite{Gu0} which was got under the assumptions:
$g_{03}=g_{13}= g_{23}=0$, the time loop lies on the "plane"
$F=\{x^0, x^3=const\}$ and $v=0$. It follows from (1.6)
that if it is true "Euclidean" relation
$$
\sigma(F) \sim \pi ^{-1}[l(L)]^2,      \eqno(1.7)
$$
where
$$
l(L)=\int\limits_{L}\sqrt{\gamma_{\alpha\beta}dx^\alpha dx^\beta}
$$
is 3-length of $L$ and $\sigma(F)$ is area of $F$, then
$$
t(L)\sim 2\cdot 10^{-24}\sqrt{\rho}\cdot [l(L)]^2 \ (sec).
$$
Hence under $\rho \sim 10^{-31} g/cm^3$  and $t\sim 1 year$ we have
$l\sim [distance \ between$ $Sun \ and \ Galaxy's \ center]
\sim 8000 parsec$;
 if $l\sim 1000 km$, then $t\sim 6\cdot 10^{-23} sec$!
When $t\sim 1 year$ and $l\sim 1000 km$, then
$\rho\sim 6\cdot 10^{28} g/cm^3$ !!

If we throw off the relation (1.7) then under
$t\sim 1 year, l\sim 1000 km$
and $\rho \sim 10^{-31} g/cm^3$  we get $\sigma\sim 10^9\pi^{-1}l^2$, i.e.
deviation from Euclidean geometry in space where exists smooth closed
time-like curve is very vast.
This means that Time machine is realized in such domains where
act strong gravitational fields which destroy the human organism.

\medskip
Suppose that functions $g_{ik}\in C^1(D)$.
It is easy to prove \cite{GuDep} that in considered domain $D$
the existence of
 smooth closed time-like curves which are homotopic to zero and are
passing
through given point $x_0\in D$ implies the equality
$$
det\mid\mid g_{ik}(x_0)\mid\mid =0,
$$
i.e. we have a singularity. Hence the smooth closed time-like
curves which
are homotopic to zero (or are contractible
to point)
 do not exist in non-singular space-time with
Euclidean topology.

\section{ The process of creation of wormholes in space-time} 

The closed time-like curve
can be constructed by means of the topological change  of domain $D$
or more
exactly by means of attaching of $4$-dimensional handle ($4$-wormhole).

How does it get  the change  of topology  in the  given
     compact domain  $D$ of space-time?  The answer is contained in
the consideration of some analogy of the Gauss-Bonnet's formula for
 closed  $3$-dimensional space-like surface $V^3 \subset D$ (the
closure  can  be got  by identification the points of border
$\partial V^3 \subset \partial D$ of non-closed $3$-surface $V^3$):
$$
\int\hspace*{-2mm}\int\limits_{V^3}\hspace*{-2mm}\int K(g_{\alpha\beta},
\frac{\partial g_{\alpha\beta}}{\partial x^{\nu}},
 \frac{\partial^2 g_{\alpha\beta}}{\partial x^\nu\partial x^\mu} )dv
= \sum_{\nu=0}^{3}c_\nu b_\nu (t), \eqno (2.1)
$$
where $K$ is the some function of $3$-metric $g_{\alpha\beta}$ and
its derivatives,
$t$ is the  time parameter, $b_{\nu}$ is the $\nu$-dimensional Betti
number,
$c_\nu$  is the real costant \cite{Gu1, Gu2, Gu3}.

The  change of  connectivity  or simple-connectivity of
3-surface  $V^3$ at the moment $t = t_0$ is realized if $b_0(t)$ or
$b_1 (t)$
are changed by jump by changing
 $t, t_0 - \delta< t < t_0  + \delta, \delta >0$.
 It doesn't take place if the curve  $L :t \rightarrow g_{\alpha\beta}(t)$
 which belongs to the space of $3$-metrics equipped with $C^2(D)$-topology
 (it means the closeness with respect to the first and second derivatives
 of $g_{\alpha\beta}$) is continuous. The discontinuity of the
curve $L$ means the existence of
 discontinuities for the second derivatives of $g_{\alpha\beta}$  on the set
 $A \subset V^3$. So the topology of $V^3$ is changed if $3$-metric
 $g_{\alpha\beta}(t) = g_{\alpha\beta}(t, x^1, x^2, x^3 )$ undergoes the
 discontinuities of the second derivatives under  $t = t_0$.

 Since $g_{\alpha\beta}(t)$ must satisfy the Einstein's equiations
then the discontinuities of $\partial_k\partial_l g_{\alpha\beta}$
take the place when
 the certain energy sources  with continuous  energy-momentum tensor
are switched on at the moment $t=t_0$ \cite{Gu1}.
The  discontinuities of  $\partial_k\partial_l g_{\alpha\beta}$
are the consequence either shock gravitational waves or another
waves the velocity of propagation of which is less than the velocity
of light.

In detail this construction of $4$-wormhole creation was analyzed in
papers \cite{Gu2, Gu3}.
 For example \cite{Gu2}, the formula (2.1) for closed orientable Riemannian
manifold $V^3$ with Killing unit vector field $\xi$ has the form
\cite{Rev}
$$
\frac{1}{2\pi l(\xi)}\int\hspace*{-2mm}\int\limits_{V^3}
\hspace*{-2mm}\int
[K(\xi_{\bot}) + 3K(\xi)]dv=2b_0(V^3) -
b_1(V^3) + d_0,
$$
where $K(\xi_{\bot})$ is Riemannian curvature in the plane that is
orthogonal
to $\xi$,  $K(\xi)$  is Riemannian curvature in the arbitrary plane
containing $\xi$, $l(\xi)$ is the length integral trajectory of
field $\xi$,
$d_0 =0$ if $b_1$ is even and $d_0=1$ if $b_1$ is odd. In the paper
\cite{Gu3}
the case
of non-compact $3$-space without any symmetries  is considered.

In the case of the General Theory of Relativity the following
general result takes the place.
If $\sigma$ is the characteristic $2$-dimensional section
of the $3$-dimensional domain $D_0$ that contains the $4$-wormhole than
the mean value
of energy density jump \cite{Gu2, Gu3}
$$
<\delta\epsilon > \sim {{c^4}\over{4\pi G}}{{1}\over{\sigma}},
$$
where $c$ is the light velocity, $G$ is the gravitational constant.

The creation of $4$-wormhole means that $3$-dimensional piece $D_0$
leaves the $3$-dimensional physical space $V^3$ or is separated from $V^3$.

\section{ Resilient space-time}         

When we have the chance to create the $4$-wormhole going from
the Present to
the Past? It is evivently if the
temporal stream carries out the peace $D_0$ to the Past which is lying
arbitrarily nearly. We can realize this by means of the theory of resilient
leaves of foliations of codimension $1$  in the
$5$-dimensional Lorentz
manifold and the conformal Kaluza-Klein theory \cite{GuSb, GuIz}.

Let $<V^4, g_{ik}>$ be a leaf of an orientable  foliation $\F$ of
codimension $1$ in
the $5$-dimensional Lorentz manifold
$<V^5, G_{AB}>, \ g=G\mid_{V^4}, \ A,B = 0,1,2,3,5$.
Foliation $\F$ is determinated by the differential 1-form
$\gamma = \gamma_Adx^A$. If the
Godbillon-Vey class $GV(\F) \neq 0$ then the foliation $\F$ has a resilient
leaves \cite{Fo}.

We suppose that  real global space-time
$V^4$ is a resilient one, i.e. is a resilient leaf of some foliation $\F$.
Hence there exists an arbitrarily small neighborhood $U_a \subset V^5$
of the event $a \in V^4$
such  that $U_a \cap V^4$ consists of at least two connected components
$U_a^1$ and $U_a^2$.

Remove the $4$-dimensional balls
$B_a\subset U_a^1, B_b\subset U_a^2$, where an event $b\in U_a^2$,
and join the boundaries
of formed two holes by means of $4$-dimensional cylinder. As result we
have a $4$-wormhole $C$, which is a Time machine if $b$ belongs
to the past of
event $a$. The past of $a$ is lying arbitrarily nearly. The distant Past
is more accessible than the near Past. A movement along 5-th coordinate
(in the direction $\gamma^A$)
gives the infinite piercing
of space-time $V^4$  at the points of Past and Future.
It is the property of a resilient
leaf \cite{Fo}.

Define the Lorentz metric $\tilde{G}_{AB}$ on $V^5$ in the following way
$$
\tilde{G}_{AB}=-\gamma_A\gamma_B + \tilde{g}_{AB},
$$
$$
  \tilde{g}_{5A} = 0,
$$
where $\tilde{g}_{AB}$ is metric tensor of $V^4$. It is more convenient
\cite{Vla}
to use conformal metric $G_{AB}$
$$
G_{AB} =\phi ^{-2}\tilde{G}_{AB}, \qquad  g_{AB} =\phi ^{-2}\tilde{g}_{AB},
\qquad \phi = \gamma_5,
$$
$$
    G _{AB}= -\lambda _A \lambda_B + g_{AB},
$$
$$
 \lambda =\phi ^{-1}\gamma ,
$$
$$
(d\tilde{I} )^2 = \tilde{G}_{AB}dx^Adx^B = \phi ^2 G_{AB}dx^Adx^B
= \phi ^2dI^2
$$
with the cylindrical condition that $G_{AB}$ are not depend of $x^5$ and
$G_{55}= -1$. Than $\phi$ is a scalar field, and the
5-dimensional Einstein's
equations
$$
R_{AB}^{(5)} - {{1}\over{2}}G_{AB}R^{(5)}= \kappa Q_{AB}
$$
are reduced \cite{Vla} to the 4-dimensional Einstein's equations, the
Maxwell's equations and the Klein-Fock equation for $\phi$
$$
R_{ik}^{(4)}- \frac{1}{2}g_{ik}R^{(4)}-\Lambda \phi^2 g_{ik} = -
{{2G}\over{c^4}}
(F_{im}F^{m\cdot}_{\cdot k} - \frac{1}{4}g_{ik}F_{mn}F^{mn})+
$$
$$
+\frac{3}{\phi }(\nabla _i \nabla _k \phi - g_{ik}g^{mn}\nabla _m
\nabla _n \phi) -
\frac{6}{\phi ^2} \phi _{,i}\phi _{,k}
+\kappa Q_{ik},                           \eqno(3.1)
$$
$$
-\nabla _m F^{mk} -3F^{mk}
\frac{\phi _{,m}}{\phi }=\frac{c^2\kappa }{\sqrt{G}}
\phi  ^3Q_A^k\lambda ^A,
$$
$$
g^{mn} \nabla _m\nabla _n \phi -\frac{1}{6}R^{(4)}\phi  +\frac{1}{3}
\Lambda \phi^3-\frac{G\phi}{2c^4} F_{mn}F^{mn} = -\frac{\kappa}{3}\phi ^3
Q_{AB}\lambda ^A\lambda ^B
$$
where $\kappa = 8\pi G/c^4, \ \phi_{,i} = \partial\phi/\partial x^i$.
The co-vector $\gamma_A$ is determinated by
the scalar field $\phi$ and 4-potential $\lambda_i, \ i=0,1,2,3$ of
electro-magnetic field $F_{ik}= \lambda_{i,k} -\lambda_{k,i}$.

Define the proper time along time-like curve $L$ as
$$
d\tau = \frac{dI}{c},\qquad dI^2 = G_{AB}dx^A dx^B,
$$
where coordinates $x^A (A=0,1,2,3,5)$ are given in domain
$U_a \subset V^5$.
Suppose that Time machine moves in the Past along $L$ in $U_a$ such that
it is at rest in the domain $V^4 \cap U_a$, i.e.
$x^1, x^2, x^3 =  const$.
Then
$$
dI^2= ds^2-d\lambda ^2,\qquad ds^2= c^2dt^2 - dl^2,
$$
where
$$
dt = \frac{g_{0i}dx^i}{c\sqrt{g_{00}}}, \qquad
dl^2=\left(-g_{\alpha \beta}+ \frac{g_{0\alpha }
g_{0\beta }}{g_{00}} \right)
dx^{\alpha }dx^{\beta } \qquad (\alpha ,\beta = 1,2,3)
$$
are chronometrical invariant respectively time and length in space-time
$V^4$. Hence
$$
d\tau  = \sqrt{1-\left(\frac{d\lambda }{ds}\right)^2 }\frac{ds}{c}=
\sqrt{1-\left(\frac{d\lambda }{ds}\right)^2 }
\sqrt{1-\left(\frac{dl}{cdt}\right)^2 }dt=
\sqrt{1-\left(\frac{d\lambda }{ds}\right)^2 }dt,   \eqno(3.2)
$$
since $dl=0$.

Let $L$ be a time-like geodesic
curve with
respect to 5-metric $G_{AB}$ that has the
ends $a$ and $b$.

Then \cite[c.51]{Vla}
$$
\frac{d\lambda}{ds} =- \frac{e}{2m\sqrt{G}},\eqno (3.3)
$$
where  $e$ is electric charge, and $m$ is mass of Time machine.

If any vector $\xi$ is tangent to $V^4$ then
$d\lambda (\xi )= \lambda _A\xi ^A=~0$. Hence the motion along curve
$L:x^A=x^A(s)$ and which is
transversal to $W^4$  is characterized by means of the inequality
$$
d\lambda (\frac{dx^A}{ds}) = \lambda _A \frac{dx^A}{ds} =
\frac{d\lambda }{ds} \neq 0. \eqno (3.4)
$$
The relations (3.3) and (3.4) imply that for transversal motion,
i.e. motion in 5-th dimension it is nesessary that body had an electric
charge. Therefore for start of Time machine we must give to it the
electric charge.

It follows from (3.2) that  $(d\lambda /ds)^2 \leq 1$, because
time $\tau$ must be real. Hence  $e/2m\sqrt{G} \leq 1$.
This inequality is not correct for electron. Thus there exists
the restriction
for mass and electric charge of Time machine.

\medskip

In the case when the Godbillon-Vey class $GV(\F) = 0$ one can attempt
to change this, i.e. to include the foliation $\F$ in smooth
one-parametric family of foliations $\F_\mu$ characteristic class
$$
Char_{\F_\mu}(\alpha)= GV(\F_\mu), \alpha \in H^3(W_1),
$$
$$
Char_{\F_\mu}(\alpha): H^*(W_1)\rightarrow H^*(V^{5}, \R)
$$
of which is changed with change of parameter $\mu$ in accord to law
$$
GV(\F_\mu)\frac{d}{d\mu}GV(\F_\mu)=0.
$$
Change (variation) is possible if $\alpha\in H^3(W_1)$ does not belong
to image of homomorphism of inclusion
$H^3(W_{2})\to H^3(W_1)$ \cite{Fu}. Since  $H^3(W_2)=0$ \cite{Fu1}
then one can be
taken abitrary cohomological class $\alpha \neq 0$.

Suppose that our space-time is not resilient one. Can it be rolled up?
It is very difficult question. As we think it has
positive answer. This is consequence of the Antropological Principle.

\section{Source of energy for Time machine}

It follows from
00-equation (3.1) that
$$
R^{(3)} + K_2 = \frac{4G}{c^4} \varepsilon _F (t) +
2\varepsilon _\phi (t) + \frac{16\pi G}{c^4} \varepsilon _Q (t).
\eqno (4.1)
$$
where $R^{(3)}$ is scalar curvature and $K_2$ is exterior one (with
respect to $V^4$) of 3-space $V^3$,
$\varepsilon _F (t),\varepsilon _\phi (t), \varepsilon _Q (t)$ are
energy densities respectively of electo-magnetic field, scalar field
and the other matter.
Hence the jumps of energy densities are consequence
of the  jump of curvature $<\delta R^{(3)}> \sim 2$ (we suppose that
$<\delta K_2>=0$),
which defines the change of topology of 3-space (see \S 1 and
\cite{Gu2, Gu3}) and separation of domain $D_0$ with
the characteristic $2$-dimensional section
$\sigma$ from space $V^3$.
The relation (4.1) shows that the jumps of
energy of electro-magnetic field and the other matter can be very vast
$$
<\delta\varepsilon _F > \sim \frac{c^4}{G}\frac{1}{\sigma},\qquad
<\delta \varepsilon _Q> \sim \frac{c^4}{4\pi G}\frac{1}{\sigma},
$$
but for scalar field $<\delta\varepsilon _\phi > \sim 1$.

\end{document}